\documentclass[11pt]{article}
\usepackage{epsfig}
\usepackage{amssymb}
\usepackage{psfrag}
\usepackage{amsmath}
\usepackage{epsfig}
\usepackage{graphicx}

\newcommand{\be}{\begin{eqnarray}}
\newcommand{\ee}{\end{eqnarray}}
\newcommand{\beq}{\begin{equation}}
\newcommand{\eeq}{\end{equation}}

\newcommand{\eq}{\begin{eqnarray}}
\newcommand{\en}{\end{eqnarray}}

\newcommand{\lsq}{{\rm l^2}}

\begin{document}

\thispagestyle{empty}

\begin{flushright}
{\footnotesize HISKP-TH-09/05, FZJ-IKP-TH-2009-17}
\end{flushright}

\vspace*{3.cm}

\begin{center}
{\Large{\bf
The role of nucleon recoil in low-energy 
antikaon-deuteron scattering
}}
\end{center}

\vspace*{0.5cm}
\begin{center}
 
{\large
V. Baru$^{a,b}$, 
E.~Epelbaum$^{a,c}$ and
 A.~Rusetsky$^c$}

\vspace*{0.5cm}

26 May 2009

\vspace*{2em}

\begin{tabular}{c}
$^a$ Forschungszentrum J\"{u}lich, Institut f\"{u}r Kernphysik (Theorie) and\\ 
J\"ulich Center for Hadron Physics,  D--52425 J\"{u}lich, Germany\\[2mm]
$^b$ Institute for Theoretical and Experimental Physics,\\
B. Cheremushkinskaya 25, 117218 Moscow, Russia\\[2mm]
$^c$ Helmholtz-Institut f\"{u}r Strahlen- und Kernphysik (Theorie) and\\ 
Bethe Center for Theoretical Physics,
Universit\"at Bonn,  D--53115 Bonn, Germany\\
\end{tabular}

\end{center}

\vspace*{2.cm}

\begin{abstract}

The effect of the nucleon recoil for  antikaon-deuteron scattering
is investigated in the framework of effective field theory. 
In particular, we concentrate on the
calculation of the nucleon recoil effect for the double-scattering
process. It is shown that the leading correction to the static
term that emerges at order  $\xi^{1/2}$  with $\xi=M_K/m_N$
vanishes due to a complete cancellation of  individually large
contributions.   The resulting recoil effect in this process
is found to be of order of 10-15{\%} as compared to the static term.
We also briefly discuss the application of the method  in the
calculations of the multiple-scattering diagrams.

\end{abstract}

\vskip1cm

{\footnotesize{\begin{tabular}{ll}
{\bf{Pacs:}}$\!\!\!\!$& 36.10.Gv, 13.75.Cs, 13.75.Jz\\
{\bf{Keywords:}}$\!\!\!\!$& Hadronic atoms, non-relativistic EFT, 
nucleon recoil
\end{tabular}}
}
\clearpage

\vspace*{-0.5cm}
\section{Introduction}
\label{sec:intro}

Recent combined analyses of pionic hydrogen and pionic deuterium
have revealed a significant progress towards a precise 
extraction of the pion-nucleon interaction parameters at threshold, see
Ref.~\cite{pid} and also~\cite{physrep,annual} for review articles.  
One might expect that the properties of the $\bar K N$ interaction
can be investigated in a similar way using a combined analysis of kaonic hydrogen 
and kaonic deuterium. The data on this sort of bound systems are provided by
the ongoing experiment of the DEAR/SIDDHARTA collaboration, which
aims to measure the $1s$ energy level shift and width of kaonic hydrogen and
kaonic deuterium, eventually with an accuracy of several eV
\cite{Dear}. Contrary to the $\pi N$ case,  the $\bar K N$
scattering length, however, is known to be large (of order of 1 fm) and strongly
absorptive. This means that in this case one will need to measure 4 independent
quantities (the real and imaginary parts of the S-wave $\bar KN$ scattering
lengths $b_0,\, b_1$), whereas
for $\pi N$ scattering one has to deal with two real scattering
lengths. Consequently, the role of the deuterium measurements is rather
different in these two cases. While the pion-nucleon scattering lengths can,
at least in principle, be determined from the pionic hydrogen alone with 
the data from pionic deuterium serving as an additional consistency check,
both data from the kaonic hydrogen and kaonic deuterium 
are needed to determine the values of $b_0,b_1$.

The analysis of the experimental data proceeds as follows. From the measured
values of the energy shift and width of the ground state in the kaonic 
hydrogen (kaonic deuterium) one first 
extracts the threshold scattering amplitudes
of $K^-p$ ($K^-d$) scattering by using the DGBT-type formulae~\cite{Deser}
at next-to-leading order in isospin breaking (see, e.g.,~\cite{physrep,MRR1,MRR2}) 
\be
 {\Delta E_{1s}} - i\, \frac{ \Gamma_{1s}}{2} &=&
 -2 \alpha^3 \mu^2 { a_p} (1-2 \mu\alpha (\ln{\alpha}-1){ a_p})\,,
\nonumber \\[2mm]
 {\Delta E^d_{1s}} - i\, \frac{ \Gamma^d_{1s}}{2} &=&
-2 \alpha^3 \mu_d^2 {  A_{\bar Kd}} \biggl(1-2 \mu_d\alpha
(\ln{\alpha}-1){  A_{\bar Kd}}\biggr), 
\ee 
where $\mu~(\mu_d)$
stands for the reduced mass of the $\bar K N$ ($\bar K d$) system and
$a_p$ ($A_{\bar Kd}$) refers to the  pertinent threshold 
amplitudes\footnote{In the case of kaonic atoms, 
higher-order Coulomb corrections may turn out to be not completely negligible 
numerically as shown, e.g.,~in Ref.~\cite{Cieply} through the exact 
solution of the Schr\"odinger equation for the bound state. 
This issue is, however, relatively easy to cure since the large contribution
comes from an iteration of a particular diagram to all orders. Replacing the factor $1-2 \mu\alpha (\ln{\alpha}-1){ a_p}$ by 
$(1+2 \mu\alpha (\ln{\alpha}-1){ a_p})^ {-1}$ already captures the bulk of the
effect. We shall not further elaborate on this issue.}. Here and in what
follows, we adopt the notation of Ref.~\cite{physrep}.

In the next step, one has to express the quantities $a_p$ and $A_{\bar Kd}$ in
terms of the S-wave $\bar KN$ scattering lengths $b_0,\, b_1$ which are defined in the
isospin-symmetric limit at $\alpha=0$ and $m_d-m_u=0$. The isospin structure
of the $\bar KN$ scattering amplitude in this world is
proportional to $b_0 + b_1 \vec \tau_K \cdot \vec\tau_N$ and the relation to the scattering 
lengths $a_0,a_1$ corresponding to the total isospin $I=0,1$
is given by $a_0=b_0-3b_1$, $a_1=b_0+b_1$. Our ultimate goal is to extract
the precise values of $b_0,b_1$ from the experiment which 
are then to be confronted with the theoretical predictions obtained
in the unitarized ChPT~\cite{Borasoy}. This would provide a beautiful test of
our knowledge about the $SU(3)$ meson-baryon dynamics at low energy.
Eventually, it will be very interesting
 to compare the experimental results
with the scattering lengths directly extracted from lattice QCD with the use of
the recently proposed method~\cite{Lage}.    

In the isospin-symmetric world, $a_p=b_0-b_1$. This relation is 
strongly modified when isospin is broken due to the unitary cusp 
effect~\cite{MRR1,Tuan}. However, the corrections due to the cusp effect
can be resummed to all orders in perturbation theory leading to the
modified expression  
for $a_p$ which is still given in terms of $b_0,b_1$ and the physical masses
of the kaons and nucleons~\cite{MRR1,Tuan}. Thus, the presence of the
cusp effect in $a_p$ does not affect the accuracy of the extraction of 
$b_0,b_1$  from the data.

The situation is rather different for the kaonic deuterium,
even in the absence of isospin breaking. In order to perform the analysis 
of the data,
the quantity $A_{\bar Kd}$ should be expressed in terms of $b_0, \, b_1$ in some kind of
multiple-scattering expansion. The procedure is plagued by our poor 
knowledge of the 3-body $\bar Kd$ interactions. It is not clear a priori  
whether the corresponding uncertainties are large enough to preclude one from
being able to extract the values of $b_0, \, b_1$ from experimental data if
such precise data would be available. A systematic model-independent analysis of the 
uncertainties in the 3-body sector is therefore needed in order to properly
analyze the kaonic deuterium data, which will be provided by SIDDHARTA
collaboration in the near future.  

The aim of the present work is to address the recoil effect, which is one
of the major sources of the theoretical uncertainty 
in the study of $\bar Kd$
scattering at low energy. Our paper is organized as follows. 
In section~\ref{sec:framework} we briefly review the theoretical framework,
which will be used to systematically investigate recoil effect in the $\bar Kd$ scattering.
The background information about the recoil effect is given in section~\ref{sec:history}.
Section~\ref{sec:double} is devoted to the discussion of the recoil
effect in the double-scattering diagram and contains both the analytic 
calculations and numerical results. Recoil effect in the multiple-scattering 
diagrams is briefly discussed in section~\ref{sec:multiple}. Finally,
our conclusions are presented in section~\ref{sec:concl}.

\section{The framework}
\label{sec:framework}

During the last few decades, the $\bar Kd$ scattering at low energy 
has been addressed on numerous occasions within the framework of Faddeev 
equations, see e.g. Refs.\cite{schick,toker,deloff,mizutani} for some of those works. 
It should, however, be pointed out that the results of these 
beautiful calculations are of no direct use in the analysis of the SIDDHARTA
data since these calculations do not provide
an explicit relation between $A_{\bar Kd}$ and $b_0, \, b_1$, which is needed for the
analysis.

Contrary to the brute-force Faddeev calculations, 
the multiple-scattering series for the $\bar Kd$
scattering length, see e.~g.~\cite{MRR2,Kamalov,Chand},
can be utilized for the analysis of the data. However,
as it is well known, the $\bar KN$ scattering lengths are large due to the 
presence of the subthreshold $\Lambda(1405)$ resonance, and the 
multiple-scattering series does not converge. The resummation of the series
can be carried out analytically using the so-called fixed center approximation
(FCA) in which nucleons are considered as static sources,
i.e.~$m_N\to\infty$. However, since 
$M_K/m_N\simeq 0.5$, one may {\it a priori} expect large corrections
to the static limit. To the best of our knowledge, no systematic method
to evaluate these {\em recoil corrections} exists in the
case of a large scattering length, where the multiple-scattering series should be resummed.
The present paper aims at the formulation of the framework, which
is capable to address this problem.

A comparison of the exact numerical solutions of the Faddeev equations for  
potential models with the multiple-scattering series resummed in the FCA 
reveals a pretty good agreement in most cases, see \cite{Gal} and references
therein. This observation provides motivation for our approach and serves
as a starting point since it indicates
that the recoil corrections could be rather small even at $M_K/m_N\simeq 0.5$
and might be amenable to the perturbative treatment. Note that
the perturbative treatment of the recoil corrections
becomes indispensable if the resummed multiple-scattering series is considered.
The potential model is a useful testing ground for our framework,
since it allows to examine the convergence of the perturbative series towards
the exactly known result.  

As  pointed out in Ref.~\cite{MRR2}, the non-relativistic
effective field theory (EFT) provides an ideal tool to explore the 
multiple-scattering expansion. In particular, one readily reproduces 
the known result~\cite{Kamalov,Chand} for static nucleons, and the inclusion
of the recoil effect is straightforward (at least, formally). Prior to going
to the details of the calculation we find it appropriate to 
discuss the essential features of the EFT approach
to the problem of interest.

\begin{itemize}

\item[i)] 
The usefulness of the multiple-scattering expansion for the 
$\bar Kd$ scattering is due to the appearance of two
distinct momentum scales. Whereas the $NN$ interactions and 3-body $\bar KNN$
interactions are mediated at large distances by 1-pion exchange, the
dominant long-distance contribution to the $\bar KN$ scattering is governed by
2-pion exchange. For this reason, we may treat $\bar KN$ interactions as
point-like, whereas $NN$ and $\bar KNN$ interactions are described
by non-local potentials.

\item[ii)]
The fact that $\Lambda(1405)$ resonance is located close to the 
$\bar KN$ threshold can, potentially, lead to enhancement of the non-local
contributions in the $\bar KN$ scattering amplitudes. 
One may expect, however,   
that the non-local effects could be still taken into 
account perturbatively, by using the effective-range expansion of the $\bar KN$
amplitude. This assumption can be checked {\it a posteriori} by explicit calculations.

\item[iii)]
The expansion parameter associated with the $\bar KN$ interaction is
given by $b\cdot\langle  1/r \rangle\simeq 1$, where $b$ denotes the
S-wave $\bar KN$ scattering length and  
$\langle  1/r \rangle\simeq 0.5~\mbox{fm}^{-1}$ is 
the expectation value of the operator $r^{-1}$ between the deuteron wave functions.
Since the expansion parameter is large, the mul\-tip\-le-scattering series does
not converge and should be resummed to all orders.

\item[iv)]
At the energies which are relevant for the problem in question, effects due to
explicit creation/annihilation of the particles can be neglected. These
processes are taken into account implicitly through various low-energy 
constants and non-local interactions present in
the Lagrangian. The non-relativistic EFT, which we use, conserves the hadron
number explicitly. 
	
\item[v)]
In addition, we expect that there is no need to include hyperonic channels $\bar KNN - \pi YN$ with
$Y=\Lambda,\Sigma$ explicitly in the non-relativistic EFT framework.
This conjecture is backed by the numerical coupled-channel
Faddeev calculations~\cite{schick,toker,mizutani,Gal}.

\item[vi)]
The (non-local) $NN$ potential can be directly imported from chiral effective
field theories, see, e.g.,~\cite{Bedaque:2002mn,Epelbaum,Epelbaum:2008ga} for
recent review articles.  The $NN$
and $\bar KN$ sectors of the non-relativistic EFT do not talk to each other.
However, in this work for demonstrative purposes we shall use NN potential in the 
separable form.

\item[vii)]
Inclusion of the three-body force is required in EFT for the consistency 
reasons alone. In order to estimate the numerical strength of the three-body force
notice that the total two-nucleon absorption rate in $K^-d$ scattering amounts
for $(1.22\pm 0.09)\%$~\cite{Veirs}. Assuming that the dispersive and absorptive 
parts of the 3-body force have the same order of magnitude, we may expect effects of the 3-body 
contribution to $ A_{\bar Kd}$ at a few percent level. 
This is definitely beyond the theoretical accuracy of the present
calculation. For comparison, note that the imaginary part of the $\pi d$ 
scattering length, which is dominated by the contribution from $\pi d\to NN$
breakup reaction, amounts more than 20\% of the real part~\cite{Hauser}.  

\item[viii)]
As we shall see below, the use of the EFT framework enables one to 
systematically obtain the expansion of $A_{\bar Kd}$ in powers of the inverse nucleon
mass. The leading-order term corresponds to the static nucleon limit 
with $m_N\to\infty$. The key assumption is that this expansion is perturbative, 
even if the multiple-scattering series is not.

\item[ix)]
In order to simplify the bookkeeping, we neglect relativistic effects for the
time being. Stated differently, we consider EFT of the underlying fundamental 
theory which is assumed to be the non-relativistic potential model. All
general conclusions should remain in place for such a model. A systematic
inclusion of the relativistic corrections will be discussed elsewhere. 
In the non-relativistic case at hand one may introduce 
the dimensionless parameter $\xi=M_K/m_N$ and consider the expansion of the
amplitude $A_{\bar Kd}$ in this parameter. Both integer and half-integer powers of
$\xi$ appear in this expansion, emerging from different momentum scales.

\item[x)]
For the time being, we completely neglect isospin-breaking effects in $\bar Kd$
scattering. They will be considered in a later publication. 
\end{itemize}

Equipped with this general clue, we now consider the 
perturbative calculation of the recoil corrections in detail. 

\begin{figure}[t]
\begin{center}
{\epsfig{file=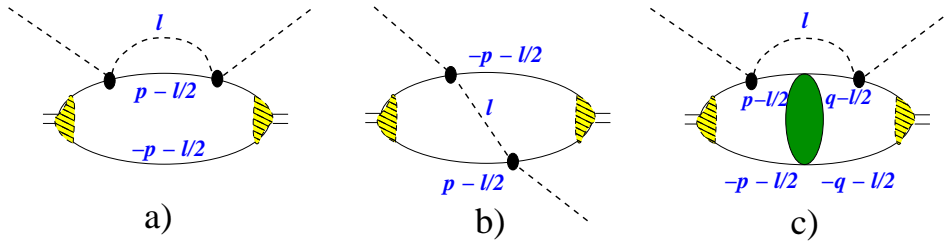, height=3.2cm}}
\end{center}
\caption{Diagrams corresponding to the meson double scattering 
on the nucleons in the deuteron}

\vspace*{.5cm}
\label{double}       % Give a unique label
\end{figure}

\section{Some known facts about nucleon recoil}
\label{sec:history}

Historically, the appearance of sizable cancellations in the case of
the $\pi d$ scattering length was   first pointed out in the papers
by Kolybasov et al. \cite{kolyb} and, independently, by F\"aldt
\cite{Faeldt} who argued that the naive static term provides
a good approximation of rescattering effects. Recently, the role
of the recoil effects was investigated for $\pi d$ scattering and
for the reaction $\gamma d\to \pi^+ nn$ within EFT
\cite{Baru_rec,Lensky_gam}. In particular, it was shown that the
3-body singularity that occurs in the $\pi NN$ intermediate state
is the origin of non-analytic corrections to the static term of
order $\sqrt{M_{\pi}/m_N}$. It was also found that the importance 
of the recoil effect is directly connected with
Pauli selection rules for the intermediate $NN$ state. In the case when
the S-wave $NN$ interaction is forbidden by the Pauli principle, the
leading $\sqrt{M_{\pi}/m_N}$ correction cancels in the diagrams of
double $\pi N$ scattering, diagrams a) and b) in
Fig.~\ref{double}, thus resulting in the small net contribution.
The situation  is different in case when the S-wave $NN$ interaction
is allowed, see diagram c) in Fig.~\ref{double}. First, the leading
$\sqrt{M_{\pi}/m_N}$  correction does not vanish here (the contributions from
diagrams a) and b) appear with the same sign). For this reason, it was
concluded in  \cite{Baru_rec,Lensky_gam} that recoil effects
should be significant in this case. Secondly, the additional
diagram with the S-wave $NN$ interaction in the intermediate state
must also be taken into account in the calculation. In the process
$\gamma d\to \pi^+ nn$, the intermediate $NN$ interaction appears to be in the
$^1S_0$ partial wave. 
Since  $NN$ interaction in the $^1S_0$ partial
wave differs considerably to the one  in the $^3S_1$ channel
there is no {\it a priori} reason to expect any kind of
cancellation between the recoil correction from diagrams similar
to a) and b) and the contribution of diagram c).
However, for $\pi d$ and $K d$ scattering the intermediate and
final $NN$ interaction occurs in the same channel ($^3S_1 -
^3\!D_1$). Therefore, a combined consideration of all diagrams of
Fig.1 is needed to draw a definite conclusion about the recoil effect\footnote{We 
neglect the D-wave contribution in the following. This does not affect the
structure of the recoil contribution, which is of our primary interest here.}. 
Note that the
Pauli-allowed recoil correction for  $\pi d$ scattering gets multiplied
with the isoscalar $\pi N$ scattering length squared which renders
this effect negligible. This is, however, not the case for
$\bar K d$ scattering where both the isoscalar and isovector
interactions are of a similar (and large) size. 

\section{$\bar K d$ scattering: Recoil effect in the double-scattering process}
\label{sec:double}

\subsection{The method}
Despite the well-known fact that
the multiple-scattering series for the $\bar K d$ scattering
requires resummation to all orders, at the first stage of our investigation we concentrate
on the double-scattering diagram. The aim is to demonstrate the technique
used to obtain a systematic expansion of any Feynman diagram in
(generally non-integer) powers of the parameter $\xi$. To this end,
we apply the
perturbative uniform expansion method proposed in Ref.~\cite{Mohr}.  
The same results can be also obtained using the threshold 
expansion method for the Feynman diagrams developed 
in Ref.~\cite{Beneke}. The advantage of the scheme of Ref.~\cite{Mohr} 
is, however, that it is not tied to a particular (dimensional) regularization.

Application of this method to carry out the integral in an arbitrary one-loop 
Feynman diagram\footnote{The method is, however, not restricted to one-loop
  diagrams.} can be summarized as follows:
\begin{itemize}
\item[i)]
Identify  the relevant momentum scales and separate the range of
 integration into the regions according to these
 scales. Suppose, for instance, that there are two distinct 
scales: the small scale $\eta$
 and the large scale $\Lambda$, $\eta/\Lambda\ll 1$. Then,
the integrand $f(\eta,q,\Lambda)$ with $q$ referring to the integration
momentum has three relevant regimes: 
\begin{itemize}
\item[]1.\,  
The low-momentum regime with $\eta\sim q \ll\Lambda$, 
\item[]2.\,
 The high-momentum regime with  $\eta\ll q\sim \Lambda$, 
\item[]3.\, 
The intermediate regime with  $ \eta\ll q \ll \Lambda$.
\end{itemize}
\item[ii)]
In each  region, perform the Taylor expansion of the integrand $f(\eta,q,\Lambda)$
in the corresponding small parameters. 
\item[iii)]
Then, the original integrand fulfills  the equality
\beq 
f(\eta,q,\Lambda)= f_{\rm l}(\eta,q,\Lambda)
 +f_{\rm h}(\eta,q,\Lambda)-f_{\rm i}(\eta,q,\Lambda)
\eeq
at the given order in
$\eta/\Lambda$. Here, $f_{\rm l},f_{\rm h},f_{\rm i},$ refer 
to the Taylor-expanded integrand $f$ in the low-, high- and 
intermediate-momentum regimes, respectively.
In particular, the integrand $f_{\rm i}$, which represents the intermediate
region, contains infrared- and ultraviolet-divergent terms  necessary to make
the integrals over the functions  $f_{\rm l}$ and $f_{\rm i}$ finite and thus plays the
role of a regulator. 
\item[iv)]
Finally, integrate the functions $f_{\rm l},f_{\rm h},f_{\rm i}$
over the whole momentum range. Since the above combination of
these functions reproduces the original function
$f(\eta,q,\Lambda)$ at the given order in $\eta/\Lambda$, the
same combination of integrals  will yield the correct
result for the original integral up to the same order.
\end{itemize}

Let  us now apply this method to the double $\bar K N$
scattering diagrams of Fig.~\ref{double}. 
The corresponding
contribution to the $\bar Kd$ scattering length reads,  
see Ref.~\cite{Baru_rec} for more details:
\beq
A_{\bar Kd}^{\sf doubl.~scatt.}
=\frac{8\pi\mu_d M_K}{\mu^2}\,(R_a+R_b+R_c)\, ,
\eeq
where the quantities $R_a$, $R_b$ and $R_c$ are defined as follows:
\eq
R_a&=&\frac{b_0^2+3b_1^2}{2M_K}\,\int\frac{d^3{\bf p} \, d^3{\bf l}}{(2\pi)^6}\,
\Psi^2\biggl({\bf p}+\frac{\bf l}{2}\biggr) \nonumber 
\bigg(
\bigg[\frac{\lsq}{2M_K}+\frac{{\rm p}^2+\gamma^2}{m_N}
+\frac{\lsq}{4m_N} \bigg]^{-1}
- \bigg[\frac{ \lsq}{2\mu} \bigg]^{-1}\bigg)\, ,
\nonumber\\[2mm]
R_b&=&\frac{b_0^2-3b_1^2}{2M_K}\,\int\frac{d^3{\bf p} \, d^3{\bf l}}{(2\pi)^6}\,
\Psi\biggl({\bf p}+\frac{\bf l}{2}\biggr)
\Psi\biggl({\bf p}-\frac{\bf l}{2}\biggr)\nonumber 
\bigg[ \frac{\lsq}{2M_K}+\frac{{\rm p}^2+\gamma^2}{m_N}
+\frac{\lsq}{4m_N} \bigg]^{-1}\, ,
\nonumber\\[2mm]
R_c&=&\frac{ b_0^2}{4m_N^2M_K}\,
\int\frac{d^3{\bf p} \, d^3{\bf q} \, d^3{\bf l}}{(2\pi)^9} \,
\Psi\left(
{\bf p}+\frac{\bf l}{2}\right) \Psi\left({\bf q}+\frac{\bf l}{2}\right) 
M_{NN}({\bf p},{\bf q},E({\rm l}))
\nonumber\\[2mm]
&\times&
\bigg[
\frac{\lsq}{2M_K}+\frac{{\rm p}^2+\gamma^2}{m_N}
+\frac{\lsq}{4m_N}
\bigg]^{-1}
\bigg[
\frac{\lsq}{2M_K}+\frac{{\rm q}^2+\gamma^2}{m_N}
+\frac{\lsq}{4m_N}
\bigg]^{-1}
\en
where $\Psi$  denotes the deuteron wave function, $\gamma^2=m_N \varepsilon_d$ with $\varepsilon_d$
being the deuteron binding energy and the energy $E({\rm l})$ is defined as  
\beq
E({\rm l})=-\varepsilon_d-\frac{\lsq}{2M_K}-\frac{\lsq}{4m_N}\,.
\eeq 
Further, $M_{NN}$ denotes the  $NN$ amplitude. Its normalization is chosen
such that in the CM frame, the amplitude $M_{NN}$ evaluated on the energy
shell is related to the S-wave scattering
phase shift $\delta(k)$ through $M_{NN}(k,k, k^2/m_N)
=16 \pi m_N (k\, \cot\delta(k)-ik)^{-1}$.
In the quantity $R_a$, renormalization
of the one-loop $\bar KN$ scattering amplitude is carried out by performing
the subtraction at threshold. At this order, this prescription yields the
same result as dimensional regularization, but has the advantage that it is not
tied to a particular regularization scheme.

In the static limit with $\xi=M_K/m_N\to 0$, only the amplitude $R_b$ survives.
In the vicinity of the static limit, each of the amplitudes $R_i$, $i=a,b,c$
can be expanded in half-integer powers of $\xi$
\beq
R_i=R_i^{\sf stat}+\xi^{1/2}\,R_i^{(1)}+\xi\,R_i^{(2)}+\xi^{3/2}\,R_i^{(3)}
+\ldots\, .
\eeq
Our aim is to perform this expansion systematically using the perturbative
uniform expansion method \cite{Mohr}. In order to do this, it is convenient to
rewrite the amplitudes $R_i$ by showing the explicit dependence on the
parameter $\xi$ in the integrands
\beq
R_a+R_b+R_c = 
b_0^2 \, (I_{\sf st}+{I_0}+{I_{NN}}+\Delta I_{\sf st})-{3 b_1^2}\,
(I_{\sf st}-{I_1}+\Delta I_{\sf st})\, ,
\eeq
where
\be\hspace*{-1.cm}
{I}_{{0}({1})}\!\!\!\!&=&\! \!\!\!\int\! \frac{d^3{\bf p}
\, d^3{\bf l}}{(2\pi)^6} \left[\Psi^2\left({\bf p}+\frac{\bf
l}{2}\!\right) {\pm}\Psi\left({\bf p}+\frac{\bf
l}{2}\!\right) \Psi\left({\bf p}-\frac{\bf l}{2}\!\right)\right] 
\left(\frac{1}
{\displaystyle {\lsq +{\xi}\left(2({\rm p}^2+\gamma^2)+ {\lsq}/{2}\right)}}
-
\frac{1}
{\lsq(1\!+\!\xi)}\right)\!,
\label{01}
\\[4mm]
\hspace*{-0.6cm}{I_{NN}}\!\!\!\!&=&\! \!\!\! \frac{{\xi}}{m_N} \int
\frac{d^3{\bf p} \, d^3{\bf q} \, d^3{\bf l}}{(2\pi)^9} 
\frac{\Psi\left(\displaystyle{\bf p}+\frac{\displaystyle\bf l}{\displaystyle 2}\right) 
\Psi\left(\displaystyle{\bf q}+\frac{\displaystyle\bf l}{\displaystyle 2}\right)M_{NN}({\bf p},{\bf q},E({\rm l})) }
{\left[{\displaystyle {\lsq +
{\xi}\left(2({\rm p}^2+\gamma^2)+\lsq/2\right)}} \right]
\left[{\displaystyle {\lsq +
{\xi}\left(2({\rm q}^2+\gamma^2)+\lsq/2\right)}} \right]},
\label{NN}
\\[4mm]
\hspace*{-1.cm}I_{\sf st}&=& \int \frac{d^3{\bf p}\, 
d^3{\bf l}}{(2\pi)^6} \,  \Psi\left({\bf p}+\frac{\bf
l}{2}\!\right) \Psi\left({\bf p}-\frac{\bf
l}{2}\!\right) \, \frac{1}{\lsq}\, , \hspace*{1.2cm} \Delta
I_{\sf st}=\frac{-\xi}{1+\xi} I_{\sf st} \,. \label{stat} 
\ee 
Here, $I_{\sf st}$ corresponds to the FCA result
(obtained in the static limit).
The recoil corrections corresponding to the Pauli-allowed
(forbidden) S-wave $NN$ intermediate state in the diagrams a) and b) of
Fig.~\ref{double} are given by the integrals $I_0$ and
$\Delta I_{\sf st}$ ($I_1$ and $\Delta I_{\sf st}$) in  
Eqs. (\ref{01}) and (\ref{stat}).
For the Pauli-allowed $NN$ state there is also a contribution from the diagram c)
given by the integral $I_{NN}$. 

We define the recoil corrections as 
$\Delta I_1=-I_1+\Delta I_{\sf st}$ and 
$\Delta I_0={I_0}+{I_{NN}}+\Delta I_{\sf st}$ 
for  isovector  and isoscalar $\bar K N$ interactions, respectively.
Below, we demonstrate how  these corrections can be evaluated
by using the method of Ref.~\cite{Mohr}.

Consider, for instance, the integral $I_1$, which 
can be rewritten in the following form 
\beq
I_1= \frac{\xi}{1+\xi}\int \frac{d^3{\bf p}d^3{\bf l}}{(2\pi)^6} 
\,\frac{f({\bf p},{\bf l})}{\lsq},\quad\quad
f({\bf p},{\bf l})=\Phi({\bf p},{\bf l})
\frac{\lsq/2-b^2}{\lsq+\xi b^2+\xi \lsq/2}\, .
\label{fun}
\eeq
Here, $b^2=2({\rm p}^2+\gamma^2)$ and the quantity $\Phi$ denotes the
following combination of the wave functions  
\beq
\label{Phi}
\Phi({\bf p},{\bf l})=\Psi^2\left({\bf p}+\frac{\bf
l}{2}\right) -\Psi\left({\bf p}+\frac{\bf
l}{2}\right) \Psi\left({\bf p}-\frac{\bf
l}{2}\right)\, .
\eeq
There are three relevant momentum 
regimes in the integral given by Eq. (\ref{fun}):

\bigskip
\noindent
\emph{1. The low-${\rm l}$ regime.} \\
In this case, the involved momenta scale as follows: 
\beq\label{poten}
\frac{\lsq}{2M_K} \sim \frac{{\rm p}^2}{ 2m_N}
\quad\Longrightarrow\quad 
{\rm l}\sim {\sqrt{\xi}}\,{\rm p}, \ \ \
{\rm p}\sim \langle 1/r \rangle_{\sf wf}\, ,
\eeq 
where $ \langle \ldots \rangle_{\sf wf}$ denotes averaging over the deuteron
wave functions. In this regime,
$\sqrt{\xi}b \sim {\rm l}$ and ${\rm l} \ll b\sim {\rm p} $.
Consequently, there are two different expansion parameters: 
$\xi$ and $\lsq/b^2$. Note, however, that 
the term $\xi b^2$ occurring in the propagator of  Eq. (\ref{fun})
is of the same order as $\lsq$ and thus 
should be kept unexpanded. 
Expanding the wave functions in Eq.~(\ref{Phi}) 
in powers of the momentum ${\bf l}$ and using $\Phi({\bf p},{\bf 0})=0$ we get
\beq
\Phi({\bf p},{\bf l})=\frac{1}{2}\,{\rm l}_i{\rm l}_j
\nabla_{\rm l}^i\nabla_{\rm l}^j\Phi({\bf p},{\bf l})\biggr|_{{\bf l}={\bf 0}}+ \, \ldots\, .
\eeq
Note that we only keep terms even in l in the expansion of  $\Phi({\bf p,l})$
since all odd terms will vanish 
after performing angular integration in Eq.~(\ref{fun}).
Expanding the integrand $f({\bf p},{\bf l})$ and averaging over the directions
according to ${\rm l}_i{\rm l}_j\to \delta_{ij}\lsq/3$ leads to  
\beq
f_{\rm l}({\bf p},{\bf l})
=\Phi_2({\bf p}) \left (-b^2+\lsq/2+\frac{\xi b^4}{\lsq+\xi b^2}\right ) + 
\ldots\, 
\quad 
\mbox{ with }
\quad
\Phi_2({\bf p}) \equiv \frac{1}{6}\nabla_{\rm l}^2\Phi({\bf p},{\bf l}) \,
\Big|_{{\bf l}=0}\,,
\eeq
where the ellipses refer to terms of a higher order in $\xi$. 
We attach the subscript ``l'' to the expanded integrand in order to signify
the low-l regime.

\bigskip
\noindent
\emph{2. The high-{\rm l} regime.}\\
In this case, the momenta ${\rm l}$ and ${\rm p}$ scale as follows:
\beq
{\rm l}\sim {\rm p} \sim \langle 1/r \rangle_{\sf wf}
\quad\Longrightarrow\quad \sqrt{\xi}b \ll {\rm l} \sim b \,.
\eeq
This implies that the function $f({\bf p},{\bf l})$ can be safely expanded 
in powers of $\xi$: 
\beq
f_{\rm h}({\bf p},{\bf l})=\Phi({\bf p},{\bf l})
 \left (\frac{-b^2+\lsq/2}{\lsq}+\xi \frac{b^4-{\rm l}^4/4}{{\rm l}^4}\right ) + \ldots\, .
\eeq
We attach the subscript ``h'' to the expanded integrand in the
high-l regime.

\bigskip
\noindent
\emph{3. The intermediate regime.}\\ This case corresponds to 
the scaling  $\sqrt{\xi}b \ll {\rm l} \ll b$. 
The expanded integrand $f_{\rm i}$ in this regime can be obtained by using
 the heavy-baryon expansion
to the function $f_{\rm l}$ or the low-momentum expansion to the function  $f_{\rm h}$. 
Both expansions lead, of course, to the same result
\beq
f_{\rm i}({\bf p},{\bf l})=\Phi_2({\bf p}) 
\left (-b^2+\lsq/2+\xi\,\frac{ b^4}{\lsq}\right ) + \ldots \,.
\eeq

\bigskip
Adding up the contributions from all three different regimes we obtain 
\beq
f_{\rm l} + f_{\rm h} -f_{\rm i} 
=\Phi({\bf p},{\bf l})  \frac{\lsq/2-b^2}{\lsq}
+\xi  \left(\Phi({\bf p},{\bf l}) \frac{b^4-{\rm l}^4/4}{{\rm l}^4} 
-\Phi_2({\bf p}) \frac{b^4}{\lsq}\right ) 
+ \xi\, \frac{\Phi_2({\bf p}) b^4}{\lsq+\xi\, b^2} +\ldots \,.
\label{funap}
\eeq 
It is now easy to identify the powers of $\xi$ which emerge after
carrying out the integration. 
The first two terms are polynomials in $\xi$. Recalling that the whole
integral is multiplied by a factor $\xi/(1+\xi)$, it is seen
that these terms start to contribute at $\mathcal{O}(\xi)$ and $\mathcal{O}(\xi^2)$, respectively.
Rescaling the integration momentum ${\rm l}\to\sqrt{\xi}\,{\rm l}$ in the third term,
one sees that this term contributes at order $\mathcal{O}(\xi^{3/2})$. 

It is easy to verify that the method of Ref.~\cite{Mohr} indeed leads to a
systematic expansion in $\xi$ by noting that the neglected 
terms $\Delta f \equiv f-(f_{\rm l}+f_{\rm h}-f_{\rm i})$,  
\beq
\Delta f=-\Phi({\bf p},{\bf l})\, 
\frac{(b^4-{\rm l}^4/4)\,(b^2+\lsq/2)}{{\rm l}^4\,(\lsq+\xi\, b^2+\xi\, \lsq/2)}\,\xi^2+
\Phi_2({\bf p}) \frac{b^6}{(\lsq+\xi\, b^2)\, \lsq}\,\xi^2\, ,
\label{diff}
\eeq
yield contributions to $I_1$ in Eq.~(\ref{fun}) of order of $\mathcal{O}(\xi^3)$ after
evaluating the corresponding integrals. 
At first sight, one might expect that rescaling ${\rm l}\to\sqrt{\xi}\,{\rm l}$ in the second term
effectively lowers the order in $\xi$, to which this term contributes
(naively, to order $\xi^{3/2}$). This is, however, not the case since 
the leading contribution is canceled by a similar one arising from the first
term in Eq.~(\ref{diff}). Therefore, Eq.~(\ref{funap}) generates all
terms in the expansion of the 
integral $I_1$ up to and including $\mathcal{O}(\xi^2)$.

For the sake of completeness, we list below terms in the expansion of 
$f({\bf p},{\bf l})$ in powers of $\xi$,
which are responsible for the contributions of orders $\xi^{5/2}$ and $\xi^3$ to the
expansion of $I_1$:
\be
\hspace*{-.5cm} f_{\rm l}+f_{\rm h}-f_{\rm i} &=& \ldots +
\xi^2\, \Phi({\bf p},{\bf l})  
\left(\frac{1}{8}-\frac{b^6}{{\rm l}^6}-\frac{b^4}{2{\rm l}^4}+\frac{b^2}{4\lsq}\right ) 
+\xi^2 \Phi_2({\bf p})\left(\frac{b^6}{{\rm l}^4}+ \frac{b^4}{2\lsq}\right )
+ \xi^2 \Phi_4({\bf p})\frac{b^6}{\lsq}
\nonumber\\[2mm]
&-&\frac{\xi^2\Phi_2({\bf p}) b^4}{2(\lsq+\xi b^2)} 
-  \frac{\xi^2\Phi_4({\bf p}) b^6}{\lsq+\xi b^2}
+  \frac{\xi^3\Phi_2({\bf p}) b^6}{2(\lsq+\xi b^2)^2}\,,
\label{funksi2}
\ee 
where ellipses refer to terms already given in Eq.~(\ref{funap}) and the
quantity $\Phi_4({\bf p})$ is defined as 
\beq
\Phi_4({\bf
  p})=\frac{1}{4!}\,\frac{1}{5}\,\nabla_{\rm l}^2\nabla_{\rm l}^2\Phi({\bf p},{\bf 
  l})\big|_{{\bf l}={\bf 0}}\,. 
\eeq

Higher-order contributions in $\xi$ can be systematically obtained along the
same lines. In particular, one verifies that the low-l regime yields only
terms with half-integer powers of $\xi$, whereas integer power terms occur
from the high-l regime. The intermediate region plays a role of
regularization leading to scale-less integrals which cancel the 
ultraviolet-divergent (infrared-divergent) terms in the low-l (high-l)
regimes.

\subsection{Results}

We are now in the position to apply the expansion method described in the previous subsection. 
In particular, one immediately observes that there is no recoil correction  at order ${\xi^{1/2}}$. 
As seen in the previous subsection, 
non-integer powers of $\xi$  can only appear from the expansion 
of the integrands in the low-l regime, see Eq. (\ref{poten}).  First, we note that
the correction $\Delta I_{\sf st}$ given by Eq. (\ref{stat}) 
does not yield the non-integer powers in $\xi$. 
Thus, performing  the expansion in  Eq.~(\ref{01}), the recoil correction for
the isovector $\bar K N$ case at order ${\xi^{1/2}}$ reads\footnote{The
  integral over l yields a contribution  $\sim 1/\sqrt{\xi}$ leading to 
the correction  for $I_{0,1}$ at order ${\xi}^{1/2}$.} 
\be
\hspace*{-.4cm}
\Delta I_1  =-{I}_{1}= 2 \xi \int \frac{d^3{\bf p}
}{(2\pi)^3} \left[\Psi^2\left({\bf p}\right)\right. {-}
\left.\Psi^2\left({\bf p}\right)\right]({\rm p}^2+\gamma^2) \int \frac{
d^3{\bf l}}{(2\pi)^3} \frac{1}{\displaystyle  \lsq}
\frac{1}{\displaystyle  {\lsq + 2{\xi}({\rm p}^2+\gamma^2))}}=0
\ee 
For the isoscalar case, the integral $I_{NN}$ contributes at the same order as $I_0$. 
At order ${\xi^{1/2}}$, the recoil correction $\Delta I_0$ can be rewritten as 
\be
\label{rec0}
\Delta I_0=I_0+I_{NN}= \frac{1}{M_K}
\int \frac{d^3{\bf p}d^3{\bf q}d^3{\bf l}}{(2\pi)^6}\ \Psi\left({\bf p}\right)
\left[G_{NN}({\bf p},{\bf q}, E ({\rm l})) -\frac{\delta({\bf p}-{\bf q})}{\lsq/2M_K} \right]
\Psi\left({\bf q}\right),
\ee
where $G_{NN}$ is the full $NN$ Green function defined as
\be\label{GNN}
G_{NN}({\bf p},{\bf q},E)=\frac{\delta({\bf p}-{\bf q})}{{\rm p^2}/m_N-E-i0}+\frac{1}{4(2\pi)^3 m_N^2}
\frac{M_{NN}({\bf p},{\bf q},E)}{{(\rm p^2}/m_N-E-i0)({\rm q^2}/m_N-E-i0)} 
\label{GNN1}
\ee
and  $E ({\rm l}) =-\varepsilon_d-\frac{\lsq}{2M_K}$ at this order. On the other hand, the Green
function $G_{NN}$ can be expressed in terms of a complete  
set of the bound- and continuum-state wave functions $\Psi(p)$ and
$\Psi^{(+)}_k(p)$, respectively, which are eigenvectors of the two-nucleon Hamiltonian\footnote{Note that
the high-l regime is perturbative, whereas the low-l regime is not.
Under this we mean that, e.g., in the low-l regime
one has to consider the full $NN$ amplitude in Eq.~(\ref{GNN}) 
without expanding it in Born series -- since all terms in this expansion contribute at the same order in $\xi$. On the other hand, if we are extracting integer powers of $\xi$ in the high-l regime, the perturbative treatment of the nucleon-nucleon scattering amplitude is legitimate, see Eq.~(\ref{xi1}).  }
\be
G_{NN}({\bf p},{\bf q},E)=\frac{\Psi\left({\bf p}\right)\Psi\left({\bf q}\right)}{-\varepsilon_d-E}
+ \int {d^3{\bf k}} \frac{\Psi^{(+)}_k({\bf p})\Psi^{(+)\dagger}_k({\bf q})}{\ {\rm k^2}/m_N-E\!-\!i0}.
\label{GNN2}
\ee
The wave functions $\psi({\bf p})=\{\Psi({\bf p}); \Psi^{(+)}_k({\bf p})\}$
satisfy the Schr\"odinger equation  
\beq
\label{schr}
\left(\epsilon-\frac{{\rm p}^2}{m_N}\right) \psi({\bf p})=-\frac{1}{4 m_N^2}\int
\frac{d^3{\bf p'}}{(2\pi)^3} V({\bf p},{\bf p}')\psi({\bf p'}) \,,
\eeq
with $\epsilon=\{-\varepsilon_d;k^2/{m_N}\}$, respectively.
%The normalization for the potential 
%is chosen so that the Born series for $M_{NN}$ are given
%by $M_{NN}=V+O(V^2)$.
The normalization for the potential is chosen so that the Lippmann-Schwinger equation  for $M_{NN}$ is given
by 
\beq
M_{NN}({\bf p},{\bf q},E)=V({\bf p},{\bf q})-\frac{1}{4m_N^2}\int
\frac{d^3{\bf p'}}{(2\pi)^3} \frac{V({\bf p},{\bf p'}) 
M_{NN}({\bf p'},{\bf q},E)}{E- {{\rm p}'^2}/m_N+i0}.
\eeq

Substituting Eq.~(\ref{GNN2}) into Eq.~(\ref{rec0}) one obtains 
\be
\Delta I_0= 2
\int \frac{d^3{\bf l}d^3{\bf k} }{(2\pi)^6}\ \frac{1}{{\lsq + 2{\xi}({\rm k}^2+\gamma^2)}}  
\left|\int d^3{\bf p}\ \Psi\left({\bf p}\right)\Psi_k^{(+)\dagger}\left({\bf p}\right) \right|^2=0,
\ee
where we exploited the othogonality of the bound- and continuum-state wave
functions and made use of the normalization of $\Psi({\bf p})$ according to 
$\int d^3{\bf p}\,  \Psi^2({\bf p})=(2\pi)^3$.	
Thus, at order $\xi^{1/2}$  there is a complete cancellation of 
the recoil corrections both for isoscalar and isovector types 
of $\bar K N$ interaction.
 The origin of the cancellation for the isovector case is
explained by the Pauli principle (see Ref.~\cite{Baru_rec}),
 whereas for the isoscalar case it
is the orthogonality of the bound-state (deuteron) and continuum-state
($NN$ intermediate state) wave functions in the $^3S_1$ partial wave.
Note that for $\pi d$-scattering  similar cancellations  were observed in
Ref. \cite{Faeldt} using a potential-model approach. 

The non-vanishing recoil corrections to the static term appear 
at order $\xi$  both for isoscalar and isovector $\bar K N$ interactions.  
In order to perform an analytic study of this and higher-order terms in the
expansion  
we choose $NN$ interaction in the separable form
$V({\bf p},{\bf p}')=\lambda g({\rm p})g({\rm p'})$, where
$g({\rm p})=({{\rm p}^2+\beta^2})^{-1}$,
$ \Psi({\bf p})=N {g({\rm p})}({{\rm p}^2+\gamma^2})^{-1}$,
$N=\sqrt{8\pi \gamma\beta(\gamma+\beta)^3}$, and  $\beta=1.4$ fm$^{-1}$.

Performing the Fourier transform and making use of the Schr\"odinger equation, one
obtains the following result for the linear corrections: 
\be\label{xi1}
\Delta I_1^{\xi}&=&
\frac{\xi}{4\pi}\,\int d^3 {\bf r} \,  r  \Psi({\bf r}) (\gamma^2 - \Delta) 
\Psi({\bf r}),
\nonumber\\[2mm]
\Delta I_0^{\xi}&=&
\frac{\xi}{4\pi}\,\int d^3 {\bf r}\,  r  \Psi({\bf r}) (\gamma^2 - \Delta) 
\Psi({\bf r})
-\frac{\xi}{16\pi m_N}
\int d^3 {\bf r} d^3 {\bf r}'   
\Psi({\bf r})\Psi({\bf r}') V({\bf r},{\bf r}')|{\bf r}-{\bf r}'|\, .
\ee
Evaluating these terms for the employed $NN$ interaction we get 
\beq
\frac{\Delta I_1^{\xi}}{I_{st}}\approx 0.6\; \xi \quad \mbox{and} \quad 
\frac{\Delta I_0^{\xi}}{I_{st}}\approx -0.3\; \xi \,.
\label{ksi}
\eeq
\begin{figure}[t]
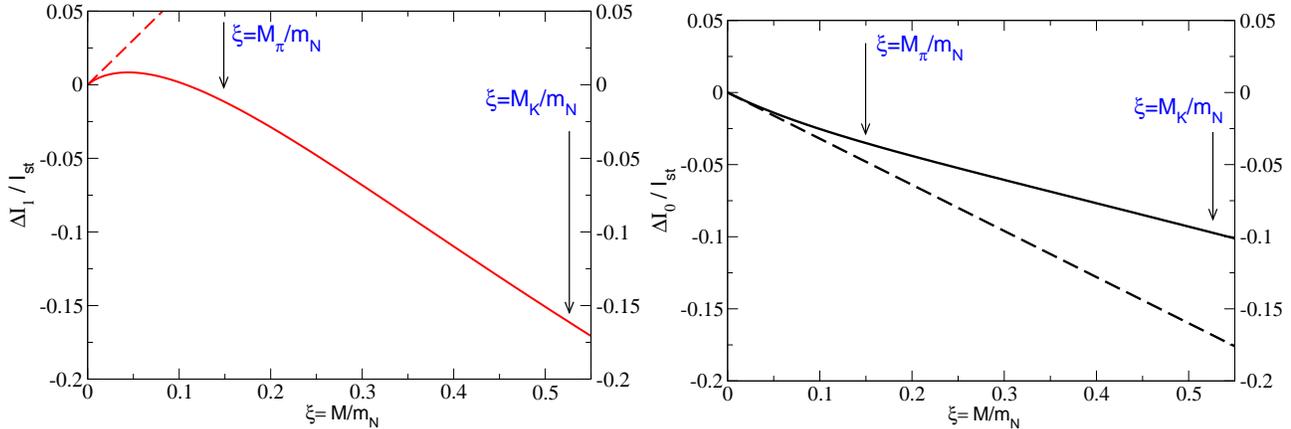

\includegraphics[width=0.495\textwidth,clip=]{rec_corrI1_ksi.eps}
\includegraphics[width=0.495\textwidth,clip=]{rec_corrI0_ksi.eps}
\caption{Recoil corrections in the double-scattering process for
the isovector (left panel) and isoscalar (right panel) cases. The solid curves correspond 
to the results of full numerical calculations (see text for more details), the dashed curves 
represent the first non-vanishing recoil corrections at order $\xi$.
The arrows indicate the results for $\pi d$- and $\bar K d$-scattering.}
\label{resrec}       % Give a unique label
\end{figure}

In Fig.~\ref{resrec} we show the results for the recoil
correction (in units of the static term $I_{\sf st}$) for isovector
(left panel) and isoscalar (right panel) contributions as a
function of $\xi$. The results of our full numerical calculation of the
recoil corrections, see Eqs. (\ref{01}), (\ref{NN}) and (\ref{stat}), 
without expanding in $\xi$ are shown by the solid curves.  
Surprisingly, even for $\bar K d$ scattering the nucleon recoil effect 
turns out to be not that large as one could {\it a priori} guess. 
As can be seen from the figure, the nucleon recoil for the double-scattering process 
amounts just to 10-15\% of the static contribution. To understand the origin
of the smallness of the effect we compare the full result with the EFT
calculation based on the expansion in powers of $\xi$. 
Our results at order $\xi$ are shown by the dashed lines.
As visualized in Fig.~\ref{resrec},  the full recoil correction changes its
sign in the considered interval of $\xi$ for the isovector case. Here, the
linear approximation fails completely to describe the full result. 
On the other hand, for the isoscalar case the recoil correction has a constant
sign, and the linear approximation yields a reasonable result.

\begin{figure}[t]
\begin{center}
\includegraphics[height=9.2cm,clip=]{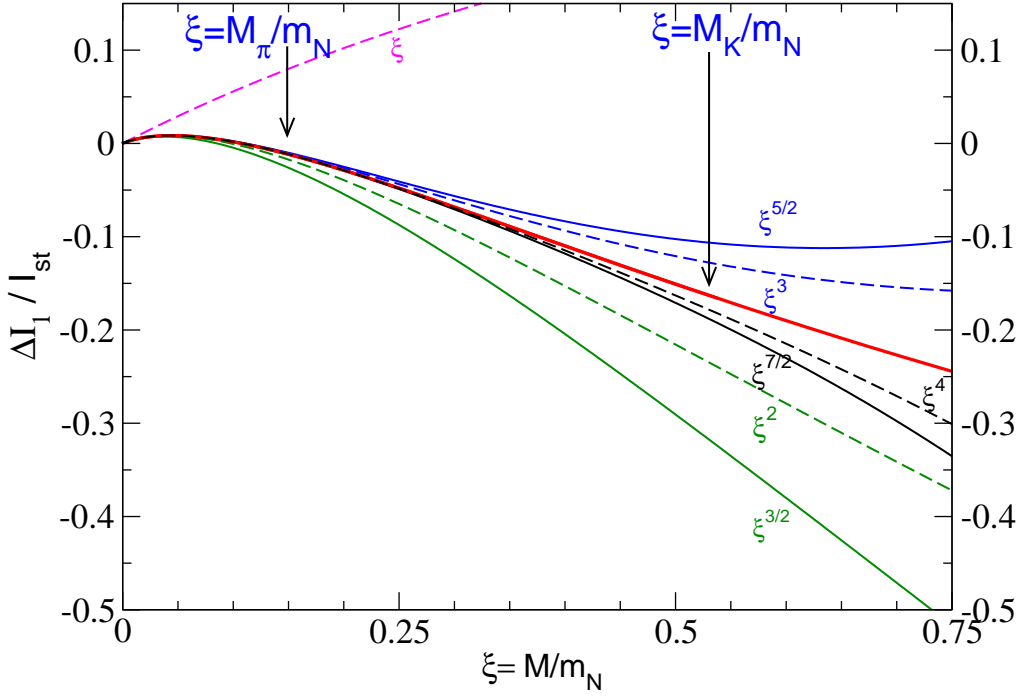}
\caption{Convergence of the expansion in $\xi$ for the recoil corrections in
  the double-scattering process for the isovector $\bar K N$ interaction.  
The notation is explained in the text. The arrows indicate the
results for $\pi d$- and $\bar K d$-scattering. Note that the (trivial)
kinematical pre-factor $1/(1+\xi)$  
was not expanded in $\xi$ (see the text for details).}
\label{converg}       % Give a unique label
\end{center}
\end{figure}
To further explore the convergence of the expansion in $\xi$ in the case of the
isovector $\bar K N$ interaction we calculated higher-order corrections  
using the expansion method described above.
The results are shown in Fig. \ref{converg}.  
The thick red line  corresponds to the full result, solid lines 
represent the calculations up to and including half-integer powers of $\xi$
whereas the results shown by dashed lines include, in addition, the
next-higher integer power of $\xi$. 
Fig.~\ref{converg} demonstrates that already at order $\xi^2$ one reproduces the
bulk of the effect while the order-$\xi^4$ calculation already provides a very good
approximation to the underlying result for $\bar K d$ scattering. 
One can see from this figure that the smallness of the net recoil effect is
accounted for by the specific 
cancellation pattern among different recoil corrections. 
In particular, there is a huge cancellation between the first integer (at order $\xi$) 
and the first non-integer (at $\xi^{3/2}$) corrections  that even leads to the
change of sign for the recoil effect.  
Further,  while improving convergence at smaller $\xi$, the inclusion of 
higher-order half-integer terms results in an oscillatory behavior around the
full result at larger $\xi$.  
Actually, looking at Eqs. (\ref{funap}) and (\ref{funksi2}), 
one can already see that the sign in front of the leading 
non-integer terms changes while going from order $\xi^{3/2}$ to
$\xi^{5/2}$ (cf. the  last term in Eq.~(\ref{funap}) and the last three terms in Eq.~(\ref{funksi2})). This can be explained by considering the
 expansion of the 3-body propagator 
in Eq. (\ref{fun}) in the low-momentum regime, which leads to an alternating
series 
\beq
f=(\lsq \Phi_2+ {\rm l}^4 \Phi_4 + \ldots)\;
\frac{\lsq/2-b^2}{\lsq+\xi b^2}\left( 1-\frac{\xi \lsq}{2(\lsq+\xi b^2)}
+\frac{\xi^2 {\rm l}^4}{4(\lsq+\xi b^2)^2}- \ldots\right)\, .
\label{fff}
\eeq
Due to this pattern, terms with half-integer (HI) powers of $\xi$ contributing to $I_1$
also show alternating behavior 
\beq
\label{I1NI}
I_1^{\rm HI}= \frac{1}{1 + \xi} (2.1 \xi^{3/2} - 0.96 \xi^{5/2} +
0.85\xi^{7/2} - 0.81 \xi^{9/2} + 0.8\xi^{11/2}
 - 0.8\xi^{13/2}
 + \ldots )\,,
\eeq
which results in their partial cancellation. 
Notice further that the (trivial)  pre-factor $1/(1+\xi)$ in
$I_1$, see Eqs.~(\ref{fun}) and~(\ref{I1NI}), produces large coefficients when expanded in
$\xi$. After expanding it in powers of $\xi$, Eq.~(\ref{I1NI}) turns into
\beq
I_1^{\rm HI} = (2.1 \xi^{3/2} - 3.07 \xi^{5/2} + 3.91\xi^{7/2} - 4.72
\xi^{9/2} + 5.53\xi^{11/2}
 - 6.33\xi^{13/2}
+ \ldots \, )\,.
\eeq
The convergence in the latter case is much slower (albeit the series still
converges at the value of $\xi$ corresponding to the kaon mass). 
For this reason, we kept this trivial kinematical pre-factor unexpanded 
when showing our results in Fig.~\ref{converg}. 

Finally, we would like to emphasize that due to large cancellations between
individually sizable terms discussed in this paragraph, 
the recoil effect turns out to be not that large even for $\bar K d$ scattering.  
We found that it  is about 10-15\% of the static 
piece for the separable model of $NN$ interaction (see arrows that indicate
the results for  $\bar K d$ scattering in Fig. \ref{resrec}).

\section{Recoil effect in multiple scattering}
\label{sec:multiple}

\begin{figure}[t]
\includegraphics[width=0.95\textwidth]{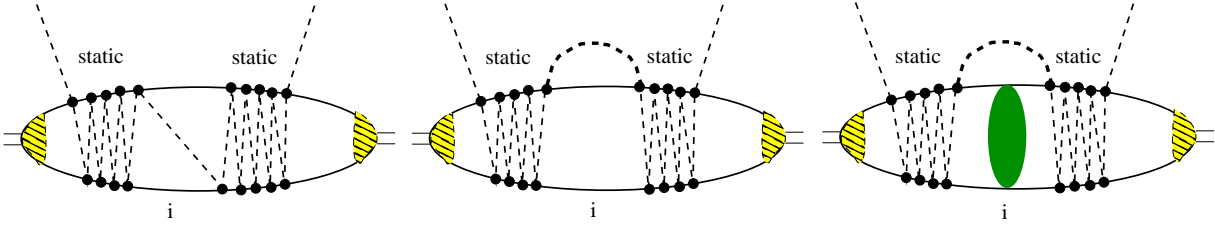}
\caption{Inclusion of the recoil corrections in the multiple-scattering process.}
\label{multiscatfig}       % Give a unique label
\end{figure}

Due to the strong $\bar K N$ interaction, the recoil
effect in the multiple-scattering diagrams might be significant and
must be studied quantitatively. Moreover, it can be shown that,
starting from the quadruple $\bar K N$ scattering process, the
recoil correction becomes nonzero already at order $\sqrt{\xi}$. A
detailed discussion of the recoil effect in the multiple scattering
will be reported elsewhere \cite{we}. Here we just sketch the
method.
The crucial assumption of the method is that the recoil corrections can be treated
perturbatively, even if the static $\bar KN$ interactions have to be resummed
to all orders. Thus, one may study an insertion of 1,2,$\ldots$ ``retarded''
blocks in the diagrams that contain infinitely many static kaon propagators.
An example is shown in Fig.~\ref{multiscatfig}, where we have dressed the
double-scattering block -- studied in detail in the previous
section -- by static kaon rescattering in the initial and final states.
In order to keep track of various powers of $\xi$ within this method, 
it is essential to have an explicit perturbative expansion of the retarded
block that can be achieved by using the technique described in the present
article.

\section{Conclusions}
\label{sec:concl}

We have studied the nucleon recoil effect for $\bar K d$
scattering using EFT. Specifically, using the expansion method of
the Feynman diagrams in EFT, we have calculated recoil corrections
to the double-scattering process in a systematic expansion in the half-integer
powers of the parameter $\xi=M_K/m_N$.
It was shown that the leading correction to the static
term, which emerges at order $\xi^{1/2}$, cancels completely both for
isoscalar and isovector types of $\bar K N$ interaction.
The origin of the cancellation for the isovector case can be
explained by the Pauli principle whereas for the isoscalar case it
stems from the orthogonality of the bound state (deuteron) and continuum
($NN$ intermediate state) wave functions in the $^3S_1$ partial
wave. The coefficients of higher order terms  in the expansion in $\xi^{1/2}$
appear to be of a natural size and the series converges even for the value
of $\xi$ corresponding to the physical kaon mass. 
Due to a significant cancellation that takes place between individually sizable terms 
at different orders, the net recoil effect for the double-scattering process is found to be just
about 10-15\% of the static contribution for the separable model of NN interaction. 
We also
sketched the method that can be used to include the nucleon recoil
for the multiple-scattering process. 
 A more detailed discussion of
this and other aspects will be presented in the forthcoming publication
\cite{we}.

%%%%%%%%%%%%%%%%%%%%%%%%%%%%%%%%%%%%%%%%%%%%%%%%%%%%%%%%%%%%%%%%%%%%%%%%%%%%%%%
\section*{Acknowledgments}

We would like to thank U.-G.~Mei{\ss}ner and A. Gal for interesting
discussions. One of the authors (A.R.) would like to thank B. Borasoy and
U. Raha for collaboration at the early stage of the project.
 The
work of E.E. and V.B. was supported in parts by funds provided
from the Helmholtz Association to the young investigator group
``Few-Nucleon Systems in Chiral Effective Field Theory'' (grant
VH-NG-222). This research is part of the EU HadronPhysics2 project ``Study of
strongly interacting matter'' under the Seventh Framework Programme of EU
(Grant agreement n. 227431). Work supported in part by DFG (SFB/TR
16, ``Subnuclear Structure of Matter''), by the 
DFG-RFBR grant (436 RUS 113/991/0-1) and by the Helmholtz
Association through funds provided to the virtual institute ``Spin
and strong QCD'' (VH-VI-231). V. B. acknowledges the support of the
Federal Agency of Atomic Research of the Russian Federation.
A.R. acknowledges the financial support of Georgia National Science Foundation
(Grant \# GNSF/ST08/4-401).

\end{document}